\documentclass[12pt,preprint]{aastex}
\usepackage{epsf}

\shorttitle{DENSITY POWER SPECTRUM}
\shortauthors{KIM \& RYU}

\begin{document}

\title{DENSITY POWER SPECTRUM OF COMPRESSIBLE HYDRODYNAMIC TURBULENT
FLOWS}

\author{Jongsoo Kim}
\affil{Korea Astronomy and Space Science Institute, Hwaam-Dong,
       Yuseong-Gu, Daejeon 305-348, Korea: {\tt jskim@kasi.re.kr}}

\and

\author{Dongsu Ryu}

\affil{Department of Astronomy and Space Science, Chungman National
       University, Daejeon 305-764, Korea: {\tt ryu@canopus.cnu.ac.kr}}

\begin{abstract}

Turbulent flows are ubiquitous in astrophysical environments, and
understanding density structures and their statistics in turbulent
media is of great importance in astrophysics.
In this paper, we study the density power spectra, $P_{\rho}$, of
transonic and supersonic turbulent flows through one and
three-dimensional simulations of driven, isothermal hydrodynamic
turbulence with root-mean-square Mach number in the range of
$1 \la M_{\rm rms} \la 10$.
From one-dimensional experiments we find that the slope of the
density power spectra becomes gradually shallower as the rms Mach
number increases.
It is because the density distribution transforms from the profile
with {\it discontinuities} having $P_{\rho} \propto k^{-2}$ for
$M_{\rm rms} \sim 1$ to the profile with {\it peaks} having
$P_{\rho} \propto k^0$ for $M_{\rm rms} \gg 1$.
We also find that the same trend is carried to three-dimension; that is,
the density power spectrum flattens as the Mach number increases.
But the density power spectrum of the flow with $M_{\rm rms} \sim 1$
has the Kolmogorov slope.
The flattening is the consequence of the dominant density structures
of {\it filaments} and {\it sheets}.
Observations have claimed different slopes of density power spectra
for electron density and cold H I gas in the interstellar medium.
We argue that while the Kolmogorov spectrum for electron density
reflects the {\it transonic} turbulence of $M_{\rm rms} \sim 1$
in the warm ionized medium, the shallower spectrum of cold H I gas
reflects the {\it supersonic} turbulence of $M_{\rm rms} \sim$ a few
in the cold neutral medium.

\end{abstract}


\section{INTRODUCTION}

According to the currently accepted paradigm, the interstellar medium
(ISM) is in a state of turbulence and the turbulence is believed to
play an important role in shaping complex structures of velocity
and density distributions \citep[see, e.g.,][for reviews]{vopg00,es04}.
The ISM turbulence is transonic or supersonic with Mach number varying
from place to place; while the turbulent Mach number, $M$, is probably
of order unity in the warm ionized medium (WIM) judging from the
temperature that ranges from $6 \times 10^3$ to $10^4$ K \citep{hrt99},
it is a few in the cold neutral medium (CNM) \citep[e.g.,][]{ht03}
and as large as 10 or higher in molecular clouds \citep[e.g.,][]{lars81}.
The fact that $M \ga 1$ implies high compressibility,
and observations of the ISM reflect the density structures resulted
from the turbulence.
Interpreting these observations and hence understanding the density
structures require a good knowledge of density statistics.

Density power spectrum, $P_{\rho}$, is a statistics that can be
directly extracted from observations.
The best-known density power spectrum of the ISM is the one
presented in \citet{acr81,ars95}.
It is a composite power spectrum of {\it electron density} collected
from observations of velocity fluctuations of the interstellar gas,
rotation measures, dispersion measures, interstellar scintillations
and others.
What is remarkable is that a single Kolmogorov slope of $-5/3$ fits
the power spectrum for the wavenumbers spanning more than 10 decades.
But there are also a number of observations that indicate
shallower spectra.
For instance, \citet{ddg00} showed that the density power spectrum
of {\it cold H I gas} has a much shallower slope of $-0.75 \sim -0.5$.
Here we note that the power spectrum of \citet{acr81,ars95} should
reveal density fluctuations preferably in the WIM since it is for
the electron density.
On the other hand, the power spectrum of \citet{ddg00} represents the
so-called tiny-scale atomic structure in the CNM \citep{heil97}.

Discussions on density power spectrum in the context of theoretical
works have been scarce.
It is partly because the study of turbulence was initiated in the
incompressible limit.
Instead velocity power spectrum was discussed extensively by
comparing the well-known theoretical spectra, such as those of
\citet{kolm41} and \citet{gs95}.
Hence, although hydrodynamic or magnetohydrodynamic (MHD)
simulations of compressible turbulence were performed,
the spectral analysis was focused mainly on velocity
(and magnetic field in MHDs) \citep[e.g.,][]{cl02,vos03}.
A few recent simulation studies, however, have reported density
power spectrum \citep[e.g.,][]{pjjn04,knp04,blc05}.
For instance, in an effort to constrain the average magnetic field
strength in molecular cloud complexes, \citet{pjjn04} have
simulated MHD turbulence of root-mean-square sonic Mach number
$M_{\rm rms} = 10$ with weak and equipartition magnetic fields.
They have found that the slope of density power spectrum is steeper
in the weak, super-Alfv\'enic case than in the equipartition case,
and super-Alfv\'enic turbulence may be more consistent with
observations.
Although their work has focused on the dependence of the slope of
density power spectrum on Alfv\'enic Mach number, they have also
demonstrated that the slope is much shallower than the Kolmogorov
slope.
\citet{knp04} and \citet{blc05} have found consistently shallow
slopes.
However, none of the above studies have yet systematically
investigated the dependency of the slope of density power spectrum
on sonic Mach number.

A note worthwhile to make is that the functional form of probability
distribution function (PDF) on the {\it density} field of compressible
turbulent flows is well established.
\citet{pv98}, by performing one-dimensional isothermal hydrodynamic
simulations, have shown that the density PDF follows a log-normal
distribution.
A few groups \citep[e.g.,][]{np99,osg01} have reported that the
log-normal distribution still holds in three-dimensional isothermal
turbulent flows even though the scaling of the standard deviation
with respect to rms Mach number is not necessarily the same as
the one obtained by \citet{pv98}.

In this paper we study the density distribution and the density
power spectrum in transonic and supersonic turbulent flows using
one and three-dimensional hydrodynamic simulations.
Specifically we show that the Mach number is an important parameter
that characterizes the density power spectrum.
The plan of the paper is as follows.
In \S 2 a brief description on numerical details is given.
Results are presented in \S 3, followed by summary and discussion
in \S 4.

\section{SIMULATIONS}

Isothermal, compressible hydrodynamic turbulence was simulated using
a code based on the total variation diminishing scheme \citep{krjh99}.
Starting from an initially uniform medium with density $\rho_o$ and
isothermal sound speed $a$ in a box of size $L$,
the turbulence was driven, following the usual procedure
\citep[e.g.,][]{macl99}; velocity perturbations were added at every
$\Delta t =0.001 L/a$, which were drawn from a Gaussian random field
determined by the top-hat power distribution in a narrow wavenumber
range of $1 \leq k \leq 2$.
The dimensionless wavenumber is defined as $k \equiv L/\lambda$, which
counts the number of waves with wavelength $\lambda$ inside $L$.
The amplitude of perturbations was fixed in such a way that $M_{\rm rms}$
of the resulting flows ranges approximately from 1 to 10.
Self-gravity was ignored.

One and three-dimensional numerical simulations were performed with
8196 and $512^3$ grid zones, respectively.
We note that one-dimensional turbulence has compressible mode only
but no rotational mode, and so its application to real situations is
limited.
However, there are two advantages: 1) high-resolution can be
achieved, resulting in a wide inertial range of power spectra and
2) the structures formed can be easily visualized and understood.
In the three-dimensional experiments the driving of turbulence was
enforced to be incompressible by removing the compressible part in
the Fourier space, although our results are not sensitive to the
details of driving.
Table 1 summarizes the model parameters.

\section{RESULTS}

Figure 1 presents the results of one-dimensional simulations.
Top two panels are snapshots of the spatial profiles of velocity
and density for a transonic turbulence with $M_{\rm rms} = 0.8$
(left) and a supersonic turbulence $M_{\rm rms} = 12.5$ (right).
Shocks are developed in both cases, but weak shocks in the transonic
case and strong shocks in the supersonic case.
The velocity profile shows shock discontinuities, superimposed
on the background that reminisces the $k=1$ and $2$ driving.
Especially, the supersonic case exhibits the so-called {\it sawtooth
profile}, which is expected in the turbulence dominated by shocks.
Like the velocity profile, the density profile for the transonic case
shows shock {\it discontinuities}.
But the density profile for the supersonic case shows high {\it peaks},
which can be understood as follows.
Since the density jump is proportional to the square of Mach number
in isothermal shocks, the density contrast at high Mach number shocks
is very large.
With the total mass conserved, the fluid should be concentrated at
shock discontinuities producing a peak-dominated profile.

Bottom two panels of Figure 1 show the time-averaged power spectra
of velocity (left) and density (right) for flows with different
$M_{\rm rms}$'s (see Table 1 for the time interval for averaging).
We note that the statistical errors (or standard deviations) of the
spectra, which are not drawn in the figure, easily exceed the averaged
values, themselves.
This is partially due to the fact that the statistical sample is small
in one-dimensional experiments. 
The slope of the velocity power spectra is nearly equal to $-2$,
irrespective of rms Mach numbers, as shown in the bottom-left panel.
On the contrary, while the slope of the density power spectrum is
close to $\sim -2$ for transonic and mildly supersonic flows, it is
much shallower for supersonic flows with $M_{\rm rms} = 7.5$ and 12.5.
Such power spectrum reflects the profiles in the top panels.
We note that discontinuities (i.e., step functions) and infinitely
thin peaks (i.e., delta-functions) in spatial distribution result in
$P_{\rho} \propto k^{-2}$ and $P_{\rho} \propto k^0$, respectively.
Hence while the density profiles with discontinuities for transonic
flows give power spectra with slopes close to $-2$, the profiles with
peaks for highly supersonic flows give shallower spectra.

Interestingly $P_{\rho} \propto k^{-2}$ and $P_{\rho} \propto k^0$
were predicted in the context of the Burgers turbulence.
It is known that the Burgers equation describes a one-dimensional
turbulence with randomly developed shocks, and the resulting sawtooth
profile gives to the $k^{-2}$ velocity power spectrum, i.e.,
$P_v \propto k^{-2}$ \citep[see, e.g.,][]{ff83}.
Expanding it, \citet{sw96} developed a model for the description
of density advected in a velocity field governed by the Burgers
equation.
They showed that in the limit of negligible pressure force (i.e.,
for strong shocks), all the mass concentrates at shock discontinuities
and the density power spectrum becomes $P_{\rho} \propto k^0$.
In the presence of pressure force (i.e., for weak shocks) their density
power spectrum is $P_{\rho} \propto k^{-2}$.
Our results agree with those of \citet{sw96}, although ours are
based on numerical simulations using full hydrodynamic equations
(under the assumption of isothermal flows).

The flattening of density power spectrum for supersonic turbulence is
also seen in three-dimensional experiments.
The results are presented in Figure 2.
Top and middle panels show the density distribution in a two-dimensional
slice for a transonic turbulence with $M_{\rm rms} = 1.2$ (top) and
a supersonic turbulence with $M_{\rm rms} = 12$ (middle).
The images reveal different morphologies for the two cases.
The transonic image includes curves of {\it discontinuities}, which
are surfaces of shocks with density jump of a few.
So the three-dimensional density distribution should contain
surfaces of discontinuities on the top of smooth turbulent background.
On the other hand, the supersonic image shows mostly density
{\it concentrations} of string and dot shapes, which are sheets and
filaments in three-dimension.
Hence, the three-dimensional density distribution should be dominated
by sheets and filaments of high density concentration.

Bottom panel of Figure 2 shows the time-averaged density power
spectra for flows with different $M_{\rm rms}$'s, along with fitted
values of slope (see Table 1 for the time interval for averaging).
As in one-dimensional experiments, the slope over the inertial range
becomes shallow as the Mach number increases.
But unlike in one-dimension, the slope for the transonic turbulence
with $M_{\rm rms} = 1.2$ is $-1.73$, close to the Kolmogorov value
of $-5/3$.
It is because the power spectrum represents the fluctuations of
turbulent background, rather than discontinuities of weak shocks.
Note that in three-dimension turbulence has the rotational mode
of eddy motions in addition to the compressional mode of sound
waves and shock discontinuities, and the normal cascade is allowed.
In supersonic turbulence, we get the slopes of $-1.08$, $-0.75$
and $-0.52$ for $M_{\rm rms} = 3.4$, $7.3$ and $12.0$, respectively.
Again the shallow spectrum reflects the highly concentrated density
distribution, but for this time, of sheet and filament morphology
shown in the middle panel.
Like delta-functions, infinitely thin sheets and filaments in
three-dimension give $P_{\rho} \propto k^0$.

As pointed in Introduction, a few recent studies of turbulence
have published the slope of density power spectrum: for instance,
\citet{pjjn04} have found a slope of $-0.71$ for $M_{\rm rms} = 10$
MHD turbulence with weak magnetic field, and \citet{knp04} have found
a slope of $-0.88$ for $M_{\rm rms} = 6$ hydrodynamic turbulence.
We notes that these values are well consistent with ours.

\section{SUMMARY AND DISCUSSION}

We present the density distribution and the density power spectrum
of driven, isothermal, compressible hydrodynamic turbulence in one
and three-dimensional numerical experiments.
The rms Mach number of flows covers from $\sim 1$ (transonic) to
$\sim 10$ (highly supersonic).
Our main findings are summarized as follows.\\
1. In transonic turbulence, the density distribution is characterized
by {\it discontinuities} generated by weak shocks on the top of
turbulent background.
On the other hand, in supersonic turbulence with $M_{\rm rms} \gg 1$,
it is characterized by {\it peaks} or {\it concentrations} of mass
generated by strong shocks.
Those density concentrations appear as sheets and filaments in
three-dimension.\\
2. In three-dimension, the slope of density power spectrum for
transonic turbulence with $M_{\rm rms} = 1.2$ is $-1.73$, which is
close to the Kolmogorov slope of $-5/3$.
But as the rms Mach number increases, the slope flattens, reflecting
the development of sheet-like and filamentary density structures.
The slopes of supersonic turbulence with $M_{\rm rms} = 3.4$, $7.3$
and $12.0$ are $-1.08$, $-0.75$ and $-0.52$, respectively.

In Introduction, we have pointed that in the ISM the power spectrum
of electron density has the Kolmogorov slope \citep{acr81,ars95},
whereas the power spectrum of H I gas shows a shallower slope
\citep{ddg00}.
Our results suggest the reconciliation of these claims of different
spectral slopes.
That is, the electron density power spectrum represents the density
fluctuations mostly in the WIM, where the turbulence is transonic and
the density power spectrum has the Kolmogorov slope.
On the other hand, the H I power spectrum represents the tiny-scale
atomic structures in the CNM, where the turbulence is supersonic with
$M_{\rm rms} \sim$ a few and the density power spectrum has a shallower
slope.

Through 21cm H I observations, it has been pointed that sheets
and filaments could be the dominated density structure in the CNM
\citep[e.g.,][]{heil97,ht03}.
Our study demonstrates that sheets and filaments are indeed the
natural morphological structures in the media with supersonic
turbulence such as the CNM.

In this work we demonstrate that the slope of density power spectrum
becomes shallow as the rms Mach number increases by considering the
simplest possible physics, i.e., by neglecting magnetic field and
self-gravity and assuming isothermality.
We point, however, that those physics could affect the
{\it quantitative} results.
For instance, the magnetic field, which is dynamically important
in the ISM, could further shallow the density power spectrum,
as \citet{pjjn04} have shown.
In addition, self-gravity could affect the power spectrum significantly,
since it forms clumps and causes the density power spectrum to become
flatter or even to have positive slopes. 
Finally, cooling could make a difference too, especially in the
turbulence in the WIM, and may even enhance compressibility,
although, as a first order approximation, the assumption of isothermality
would be adequate in molecular clouds \citep[see, e.g.,][]{psm05}.

\acknowledgments

We thank the referee, P. Padoan, for constructive comments, and
J. Cho and H. Kang for discussions.
The work by JK was supported by KOSEF through Astrophysical
Research Center for the Structure and Evolution of Cosmos (ARCSEC).
The work by DR was supported by Korea Research Foundation Grant
(KRF-2004-015-C00213). Numerical simulations were performed using
``Linux Cluster for Astronomical Computations'' of KASI-ARCSEC.

\clearpage

\begin{deluxetable}{lcllcc}
\tabletypesize{\scriptsize}
\tablecaption{Model Parameters\tablenotemark{a}}
\tablewidth{0pt}
\tablehead{
\colhead{Model\tablenotemark{b}} & 
\colhead{$\stackrel{\cdot}{E}_{\rm kin}$\tablenotemark{c}} & 
\colhead{$t_{\rm end}$\tablenotemark{d}} & 
\colhead{$dt_{\rm output}$\tablenotemark{e}} &
\colhead{$\Delta t_{\rm sp}$\tablenotemark{f}} &
\colhead{Resolution}
}
\startdata
1D0.8  & 0.1  &  8    &  0.1  & 4-8      &  8196    \\
1D1.7  & 1    &  8    &  0.1  & 4-8      &  8196    \\
1D3.4  & 10   &  8    &  0.1  & 4-8      &  8196    \\
1D7.5  & 100  &  8    &  0.1  & 4-8      &  8196    \\
1D12.5 & 400  &  8    &  0.1  & 4-8      &  8196    \\
3D1.2  &  1   &  4.8  &  0.4  & 1.2-4.8  &  $512^3$ \\
3D3.4  &  30  &  1.5  &  0.1  & 0.6-1.5  &  $512^3$ \\
3D7.3  &  300 &  1.0  &  0.05 & 0.5-1.0  &  $512^3$ \\
3D12.0 &  1300&  0.5  &  0.05 & 0.3-0.5  &  $512^3$ \\
\enddata
\tablenotetext{a}{All the quantities are given in the units of
$\rho_o$, $a$ and $L$.}
\tablenotetext{b}{1D or 3D for one or three-dimension followed by
the rms Mach number.}
\tablenotetext{c}{kinetic energy input rate}
\tablenotetext{d}{end time of each simulation}
\tablenotetext{e}{time interval for data output}
\tablenotetext{f}{time interval over which power spectra
were averaged}
\end{deluxetable}

\clearpage

\begin{figure}
\vskip -2cm \plotone{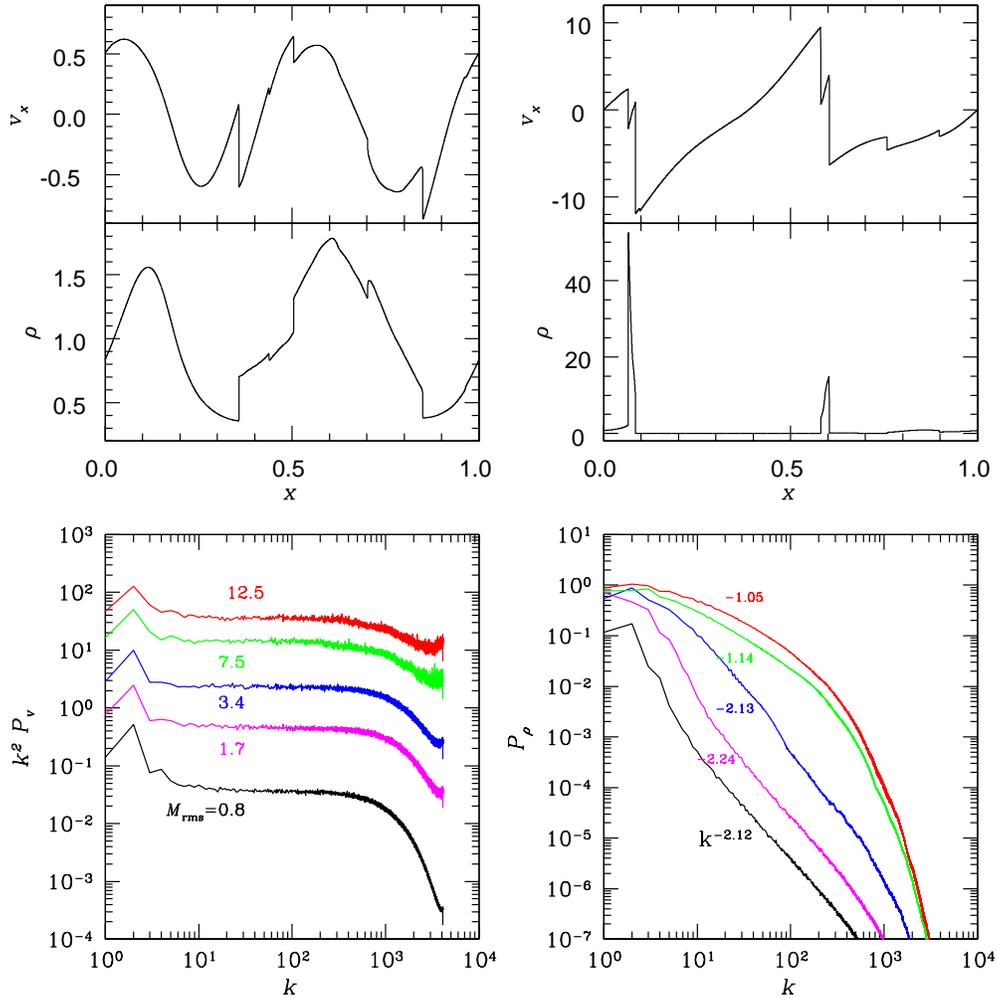} \vskip -5cm \figcaption
{One-dimensional turbulence.
{\it Top panels}: Snapshots of the spatial profiles of velocity
and density, after turbulence was fully saturated, for a transonic
turbulence with time averaged $M_{\rm rms} = 0.8$ (left) and a
supersonic turbulence with time averaged $M_{\rm rms} = 12.5$ (right).
The plotted quantities are normalized by the isothermal sound speed
and the initial density, respectively.
{\it Bottom panels}: Time-averaged power spectra of velocity (left) and
density (right) for flows with different $M_{\rm rms}$'s.
The velocity power spectrum in the left panel is multiplied by $k^2$
for clarity.
Slopes in the right panel were obtained by least-square fits over
the range of $20 \le k \le 60$.
Simulations used 8196 grids zones.}

\end{figure}

\begin{figure}
\vspace{0cm}\hspace{4.5cm}\epsfxsize=6.5cm\epsfbox{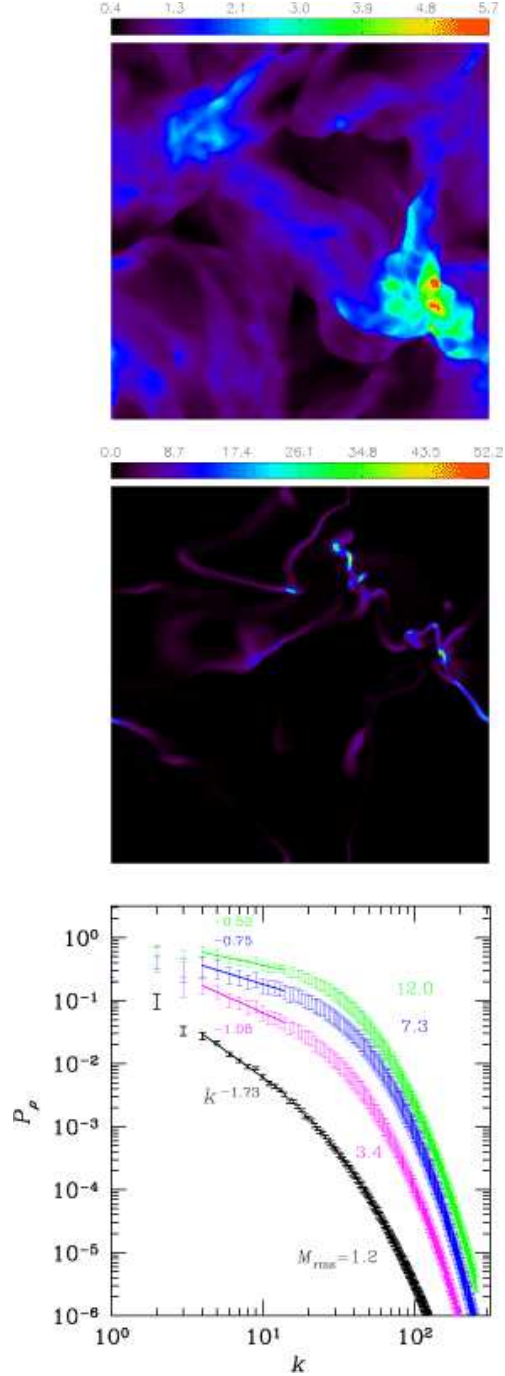}\vspace{0cm}
\figcaption
{Three-dimensional turbulence.
{\it Top and middle panels}: Color images of the density distribution
in a two-dimensional slice, after turbulence was fully saturated, for a
transonic turbulence with time averaged $M_{\rm rms} = 1.2$ (top) and
a supersonic turbulence with time averaged $M_{\rm rms} = 12$ (middle).
The color is coded on linear scales and the range of density is
an order of magnitude larger for the supersonic case than for the
transonic case.
{\it Bottom panel}: Statistical error bars of the time-averaged density
power spectra for flows with different $M_{\rm rms}$'s.
Solid lines and their slopes, which were obtained by least-square fits
over the range of $4 \le k \le 14$, are included.
Simulations used $512^3$ grids zones.}
\end{figure}

\end{document}